\documentclass[aps,floatfix,prd,preprintnumbers]{revtex4}
\usepackage{amssymb}
\usepackage{amsmath}
\usepackage{epstopdf}
\usepackage{hyperref}
\usepackage{capt-of}
\usepackage{graphicx}
\usepackage{dcolumn}
\usepackage{bm}

\RequirePackage{float}
\begin{document}

\title{Localization of Energy and Momentum in an Asymptotically
Reissner-Nordstr\"{o}m Non-singular Black Hole Space-time Geometry}
\author{Irina Radinschi}
\affiliation{Department of Physics, ``Gheorghe Asachi" Technical University, Iasi,
700050, Romania, Email: radinschi@yahoo.com}
\author{Pradyumn Kumar Sahoo}
\affiliation{Department of Mathematics, Birla Institute of Technology and Science-Pilani,
Hyderabad Campus, Hyderabad-500078, India, Email:
pksahoo@hyderabad.bits-pilani.ac.in}
\author{Theophanes Grammenos}
\affiliation{Department of Civil Engineering, University of Thessaly, 383 34 Volos,
Greece Email: thgramme@civ.uth.gr}
\author{Surajit Chattopadhyay}
\affiliation{Department of Mathematics, Amity University, West Bengal, Kolkata 700135,
India, Email: surajcha@associates.iucaa.in}
\author{Marius-Mihai Cazacu}
\affiliation{Department of Physics, \textquotedblleft Gheorghe Asachi" Technical
University, Iasi, 700050, Romania, Email: marius.cazacu@tuiasi.ro}

\begin{abstract}
\begin{center}
\textbf{Abstract}\\
\end{center}
The space-time geometry exterior to a new four-dimensional, spherically
symmetric and charged black hole solution that, through a coupling of
general relativity with a non-linear electrodynamics, is everywhere
non-singular, for small $r$ it behaves as a de Sitter metric, and
asymptotically it behaves as the Reissner-Nordstr\"{o}m metric, is
considered in order to study the energy-momentum localization. For the
calculation of the energy and momentum distributions, the Einstein,
Landau-Lifshitz, Weinberg and M\o ller energy-momentum complexes have been
applied. The results obtained show that in all prescriptions the energy
depends on the mass $M$ of the black hole, the charge $q$, two parameters $%
a\in \mathbb{Z}^+$ and $\gamma\in \mathbb{R}^+$, and on the radial
coordinate $r$. The calculations performed in each prescription show that
all the momenta vanish. Additionally, some limiting and particular cases for
$r$ and $q$ are studied, and a possible connection with strong gravitational
lensing and micro lensing is attempted.
\end{abstract}

\maketitle


\section{Introduction}

The problem of energy-momentum localization, being one of the most
challenging problems in classical general relativity in the search of a
physically meaningful expression for the energy-momentum of the
gravitational field, has triggered a lot of interesting research work, but
still remains open and rather not fully understood. As a result, there is
not a generally accepted definition for the notion of the localized
energy-momentum associated with the gravitational field. Nevertheless,
several and often different approaches have been tried in an attempt to
achieve the aforementioned energy-momentum localization. Among the
mathematical tools utilized, the most notable are superenergy tensors
[1]-[3], quasi-local energy definitions [4]-[8] and a number of the so
called energy-momentum complexes [9]-[15]. In particular, the
energy-momentum complexes of Einstein [9]-[10], Landau-Lifshitz [11],
Papapetrou [12], Bergmann-Thomson [13] and Weinberg [15] are
coordinate-dependent pseudo-tensorial quantities which can be used in
Cartesian and quasi-Cartesian coordinates, more precisely in Schwarzschild
Cartesian coordinates and in Kerr-Schild Cartesian coordinates, and have
yielded so far many physically meaningful as well as interesting results
[16]-[24]. In fact, it has been found that different pseudo-tensor complexes
lead to the same energy for any metric of the Kerr-Schild class and even for
space-times more general than those described by this class (see, e.g.,
[25], [26] and [27] for some reviews and references therein). Furthermore,
the M\o ller energy-momentum complex [14] allows the calculation of energy
and momenta in any coordinate system, including the aforesaid Schwarzschild
Cartesian coordinates and Kerr-Schild Cartesian coordinates, and it has also
provided a number of physically interesting results for many space-time
geometries, in particularly for four-dimensional, three-dimensional,
two-dimensional, and one-dimensional space-times [28]-[39].

More recently, the relevant research has also turned to the teleparallel
theory of gravitation whereby a number of similar results for the energy of
the gravitational field has been obtained [40]-[48], while one should also
underline the notion of the quasi-local mass introduced by Penrose [49] and
further developed by Tod [50] for various gravitating systems. It must be
stressed that the Einstein, Landau-Lifshitz, Papapetrou, Bergmann-Thomson,
Weinberg and M\o ller prescriptions agree with this quasi-local mass
definition. Furthermore, some rather pertinent and modern approaches concern
the quasi-local energy-momentum associated with a closed 2-surface, and the
concept of the Wang-Yau quasi-local energy [51-52]. Indeed, the effort to
rehabilitate the value of the energy-momentum complexes has led to the study
of pseudo-tensors and quasi-local approaches in the context of a Hamiltonian
formulation with a choice of a four-dimensional isometric Minkowski
reference geometry on the boundary. It was found that for any closed
2-surface there exists a common value for the quasi-local energy for all
expressions that agree (to linear order) with the Freud superpotential. In
other words, all the quasi-local expressions in a large class yield the same
energy-momentum [53], [54].

The present paper is organized as follows. In Section 2 the new
four-dimensional, non-singular and charged black hole space-time that
asymptotically behaves as the Reissner-Nordstr\"{o}m solution is described.
Then, the Einstein pseudo-tensorial complex applied for the calculation of
the energy and momentum distributions is introduced in Section 3, together
with the results obtained. In Section 4, the Landau-Lifshitz energy-momentum
complex and the calculated expressions for energy-momentum are presented.
Section 5 is devoted to the depiction of the Weinberg prescription and the
presentation of the evaluated expressions for the energy and momentum. In
Section 6 we introduce the M\o ller energy-momentum complex and show the
results obtained. Section 7 contains a discussion of the main
results and some limiting and particular cases. Finally, in Section 8 we present the implied conclusions of our study. In our calculations we have used geometrized units ($c=G=1$) and
for the signature of the metric we have chosen ($+$,$-$,$-$,$-$). We have
used the Schwarzschild Cartesian coordinates \{$t$, $x$, $y$, $z$\} in the
case of the Einstein, Landau-Lifshitz and Weinberg prescriptions, and the
Schwarzschild coordinates \{$t$, $r$, $\theta $, $\phi $\} for the M\o ller
prescription, respectively. Finally, Greek indices take values from $0$ to $%
3 $, while Latin indices run from $1$ to $3$.

\section{The Asymptotically Reissner-Nordstr\"{o}m Non-singular Black Hole
Space-time Geometry}

In this section we introduce \ the asymptotically Reissner-Nordstr\"{o}m
non-singular black hole space-time geometry. This new spherically symmetric
and charged non-singular black hole is built based on the metric function $%
f(r)=1-\frac{2M}{r}\left[ \frac{1}{1+\gamma \left( \frac{q^{2}}{Mr}\right)
^{a}}\right] ^{3/a}$ given by eq. (56) in the L. Balart and E. C. Vagenas
paper [55]. The method consists in adding a new term which will make the
metric behave asymptotically as the Reissner--Nordstr\"{o}m metric. For the
new term, the Dagum distribution function that contains a factor $\frac{q}{%
r^{2}}$ was employed, so that the metric function is given by
\begin{equation}
f(r)=1-\frac{2M}{r}\left[ \frac{1}{1+\gamma \left( \frac{q^{2}}{Mr}\right)
^{a}}\right] ^{3/a}+\frac{q^{2}}{r^{2}}\left[ \frac{1}{1+\gamma \left( \frac{%
q^{2}}{Mr}\right) ^{a}}\right] ^{4/a},  \tag{1}
\end{equation}
with $a\geq 2$ representing an integer and a constant $\gamma \in \mathbb{R}%
^+$. By setting $\gamma \geq (2/3)^{a}$ it is seen that the black hole
solution satisfies the weak energy condition. The associated electric field
is expressed as
\begin{equation}
E(r)=\frac{q}{r^{2}}\left(\frac{3\,\gamma (3+a)(\frac{q^{2}}{Mr})^{a-1}}{2\left[
(1+\gamma \left( \frac{q^{2}}{Mr}\right) ^{a}\right] ^{2+3/a}}+\frac{%
1-\gamma (3+a)\left( \frac{q^{2}}{Mr}\right) ^{a}}{\left[(1+\gamma \left( \frac{%
q^{2}}{Mr}\right) ^{a}\right]^{\frac{2(2+a)}{a}}}\right).  \tag{2}
\end{equation}

For small values of the $r$ coordinate, the black hole metrics obtained from
eq. (1) show a de Sitter black hole behaviour with

\begin{equation}
f(r)\approx 1-\frac{M^{4}}{\gamma ^{\frac{4}{a}}\,q^{6}}(2\,\gamma ^{\frac{1%
}{a}}-1)r^{2}.  \tag{3}
\end{equation}

Notice that the factor in front of the term $r^{2}$ cannot become zero
because the values of $\gamma $ are restricted at $\gamma \geq (2/3)^{a}$.
For $(1/2)^{a}\leq \gamma <(2/3)^{a}$, these black hole metrics remain
non-singular, without satisfying the weak energy condition, but they exhibit
a de Sitter center. The conclusion is that if a black hole metric is
non-singular and satisfies the weak energy condition, then it possesses a de
Sitter center, but if the metric has a de Sitter behavior when approaching
the center, this does not necessarily imply that it satisfies the weak
energy condition. For $\gamma <(1/2)^{a}$, the black hole metric is
singular. A special case of eq. (1) arises when we choose $a=2$ and $\gamma
=M^{2}/q^{2}$, corresponding to, as it was pointed out in [55], the black
hole metric presented in [56].

\bigskip Thus, the new spherically symmetric, static and charged
asymptotically Reissner-Nordstr\"{o}m non-singular black hole metric
considered is described by a line element of the form
\begin{equation}
ds^{2}=B(r)dt^{2}-A(r)dr^{2}-r^{2}(d\theta ^{2}+\sin ^{2}\theta d\phi ^{2}),
\tag{4}
\end{equation}%
with $B(r)=f(r)$, $A(r)=\frac{1}{f(r)}$, and the metric function is given by
eq. (1).

\section{Einstein Prescription and the Energy Distribution of the
Asymptotically Reissner-Nordstr\"{o}m Non-singular Black Hole}

The Einstein energy-momentum complex [9]-[10] for a ($3+1$)-dimensional
space-time has the expression
\begin{equation}
\theta _{\nu }^{\mu }=\frac{1}{16\pi }h_{\nu ,\,\lambda }^{\mu \lambda }.
\tag{5}
\end{equation}%
The superpotentials $h_{\nu }^{\mu \lambda }$ are given by
\begin{equation}
h_{\nu }^{\mu \lambda }=\frac{1}{\sqrt{-g}}g_{\nu \sigma }\left[ -g(g^{\mu
\sigma }g^{\lambda \kappa }-g^{\lambda \sigma }g^{\mu \kappa })\right]
_{,\kappa }  \tag{6}
\end{equation}%
and exhibit the property of antisymmetry
\begin{equation}
h_{\nu }^{\mu \lambda }=-h_{\nu }^{\lambda \mu }.  \tag{7}
\end{equation}%
In the Einstein prescription the local conservation law is written:
\begin{equation}
\theta _{\nu ,\,\mu }^{\mu }=0.  \tag{8}
\end{equation}%
The energy and momentum are evaluated by
\begin{equation}
P_{\mu }=\iiint \theta _{\mu }^{0}\,dx^{1}dx^{2}dx^{3},  \tag{9}
\end{equation}%
where $\theta _{0}^{0}$ and $\theta _{i}^{0}$ represent the energy and
momentum density components, respectively.

Applying Gauss' theorem, the energy-momentum has the expression
\begin{equation}
P_{\mu }=\frac{1}{16\pi }\iint h_{\mu }^{0i}n_{i}dS,  \tag{10}
\end{equation}%
with $n_{i}$ the outward unit normal vector over the surface element $dS.$
In eq. (10) $P_{0}$ is the energy.

To calculate the energy distribution and momentum, the line element given by
eq. (1) is converted to Schwarzschild Cartesian coordinates $t,x,y,z$ using
the coordinate transformation
\begin{equation}
\begin{split}
x& =r\sin \theta \cos \phi , \\
y& =r\sin \theta \sin \phi , \\
z& =r\cos \theta .
\end{split}
\tag{11}
\end{equation}%
The line element given by eq. (1) and eq. (4) reads now%
\begin{equation}
ds^{2}=f(r)dt^{2}-(dx^{2}+dy^{2}+dz^{2})-\frac{\frac{1}{f(r)}-1}{r^{2}}%
(xdx+ydy+zdz)^{2},  \tag{12}
\end{equation}

where
\begin{equation}
r=\sqrt{x^{2}+y^{2}+z^{2}}.  \tag{13}
\end{equation}

For $\mu =0,1,2,3$ and $i=1,2,3$ we conclude that the following components
of the superpotential $h_{\mu }^{0i}$ in quasi-Cartesian coordinates vanish:
\begin{equation}
\begin{split}
& h_{1}^{01}=h_{1}^{02}=h_{1}^{03}=0, \\
& h_{2}^{01}=h_{2}^{02}=h_{2}^{03}=0, \\
& h_{3}^{01}=h_{3}^{02}=h_{3}^{03}=0
\end{split}
\tag{14}
\end{equation}%
while the non-vanishing components of the superpotential are given by
\begin{equation}
h_{0}^{01}=\frac{2x}{r^{2}}\left\{\frac{2M}{r}\left[ \frac{1}{1+\gamma \left(
\frac{q^{2}}{Mr}\right) ^{a}}\right] ^{3/a}-\frac{q^{2}}{r^{2}}\left[ \frac{1%
}{1+\gamma \left( \frac{q^{2}}{Mr}\right) ^{a}}\right] ^{4/a}\right\},  \tag{15}
\end{equation}

\begin{equation}
h_{0}^{02}=\frac{2y}{r^{2}}\left\{\frac{2M}{r}\left[ \frac{1}{1+\gamma \left(
\frac{q^{2}}{Mr}\right) ^{a}}\right] ^{3/a}-\frac{q^{2}}{r^{2}}\left[ \frac{1%
}{1+\gamma \left( \frac{q^{2}}{Mr}\right) ^{a}}\right] ^{4/a}\right\},  \tag{16}
\end{equation}

\begin{equation}
h_{0}^{03}=\frac{2z}{r^{2}}\left\{\frac{2M}{r}\left[ \frac{1}{1+\gamma \left(
\frac{q^{2}}{Mr}\right) ^{a}}\right] ^{3/a}-\frac{q^{2}}{r^{2}}\left[ \frac{1%
}{1+\gamma \left( \frac{q^{2}}{Mr}\right) ^{a}}\right] ^{4/a}\right\}.  \tag{17}
\end{equation}

Combining the expression for the energy-momentum distribution given by eq.
(10) and eqs. (15)-(17) the expression for the energy distribution in the
Einstein prescription for the non-singular and charged black hole space-time
that asymptotically behaves as the Reissner-Nordstr\"{o}m solution is given
by
\begin{equation}
E_{E}=M\left[ \frac{1}{1+\gamma \left( \frac{q^{2}}{Mr}\right) ^{a}}\right]
^{3/a}-\frac{q^{2}}{2r}\left[ \frac{1}{1+\gamma \left( \frac{q^{2}}{Mr}%
\right) ^{a}}\right] ^{4/a}.  \tag{18}
\end{equation}%
From eq. (10) and eq. (14), we obtain that all the momentum components
vanish:
\begin{equation}
P_{x}=P_{y}=P_{z}=0.  \tag{19}
\end{equation}

In Fig. 1 we plot the energy in the Einstein prescription for the choice of
parameters $a=2$ and $\gamma =\frac{4}{9}$, and with $M=1$ and $q=0.1$.

\begin{figure}[H]
\begin{center}
\includegraphics[scale=1.2]{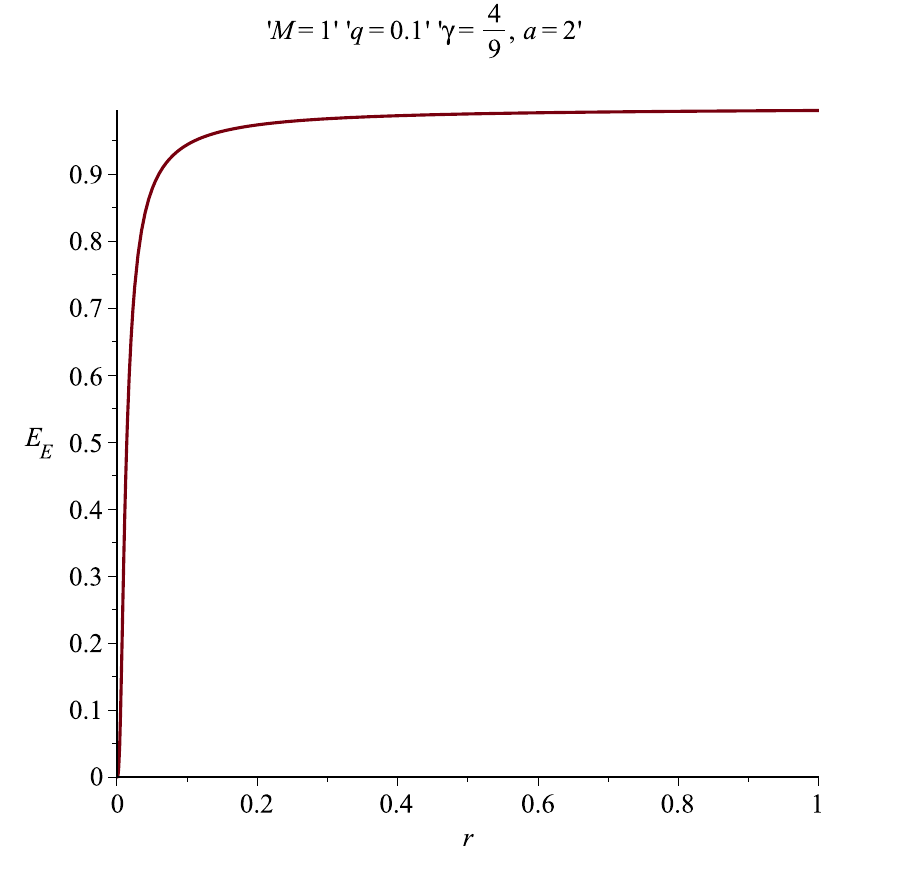}
\end{center}
\caption{Einstein energy versus the radial distance $r$.}
\label{f1}
\end{figure}

Fig. 2 shows the behaviour of the energy near the origin for the same values
of $a$, $M$, $q$ and $\gamma $.

\begin{figure}[H]
\centering
\includegraphics[scale=1.2]{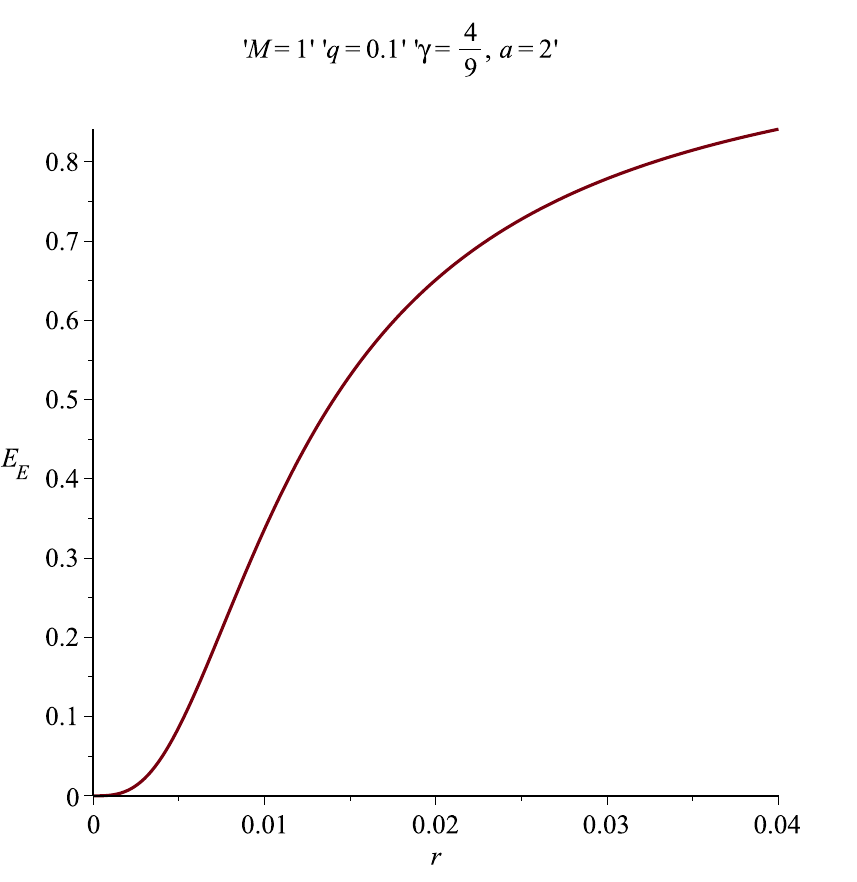}
\caption{Einstein energy versus the radial distance $r$ near the origin.}
\label{f2}
\end{figure}

\section{\protect\bigskip Landau-Lifshitz Energy-Momentum Complex and the
Energy Distribution of the Asymptotically Reissner-Nordstr\"{o}m
Non-singular Black Hole}

The Landau-Lifshitz energy-momentum complex is given by [11]
\begin{equation}
L^{\mu \nu }=\frac{1}{16\pi }S_{,\,\rho \sigma }^{\mu \nu \rho \sigma },
\tag{20}
\end{equation}%
whith the Landau-Lifshitz superpotentials
\begin{equation}
S^{\mu \nu \rho \sigma }=-g(g^{\mu \nu }g^{\rho \sigma }-g^{\mu \rho }g^{\nu
\sigma }).  \tag{21}
\end{equation}%
The $L^{00}$ and $L^{0i}$ components represent the energy and the momentum
densities, respectively. For the Landau-Lifshitz prescription the local
conservation is respected
\begin{equation}
L_{,\,\nu }^{\mu \nu }=0.  \tag{22}
\end{equation}%
By integrating $L^{\mu \nu }$ over the 3-space, one obtains the expressions
for the energy and momentum:
\begin{equation}
P^{\mu }=\iiint L^{\mu 0}\,dx^{1}dx^{2}dx^{3}.  \tag{23}
\end{equation}%
With the aid of Gauss' theorem we get
\begin{equation}
P^{\mu }=\frac{1}{16\pi }\iint S_{,\nu }^{\mu 0i\nu }n_{i}dS=\frac{1}{16\pi }%
\iint U^{\mu 0i}n_{i}dS.  \tag{24}
\end{equation}

In the Landau-Lifshitz prescription, the calculations have to be performed
using the line element (12). The non-vanishing components of the the
Landau-Lifshitz superpotentials are

\begin{equation}
U^{001}=\frac{2x}{r^{2}}\frac{\frac{2M}{r}\left[ \frac{1}{1+\gamma \left(
\frac{q^{2}}{Mr}\right) ^{a}}\right] ^{3/a}-\frac{q^{2}}{r^{2}}\left[ \frac{1%
}{1+\gamma \left( \frac{q^{2}}{Mr}\right) ^{a}}\right] ^{4/a}}{1-\frac{2M}{r}%
\left[ \frac{1}{1+\gamma \left( \frac{q^{2}}{Mr}\right) ^{a}}\right] ^{3/a}+%
\frac{q^{2}}{r^{2}}\left[ \frac{1}{1+\gamma \left( \frac{q^{2}}{Mr}\right)
^{a}}\right] ^{4/a}},  \tag{25}
\end{equation}

\begin{equation}
U^{002}=\frac{2y}{r^{2}}\frac{\frac{2M}{r}\left[ \frac{1}{1+\gamma \left(
\frac{q^{2}}{Mr}\right) ^{a}}\right] ^{3/a}-\frac{q^{2}}{r^{2}}\left[ \frac{1%
}{1+\gamma \left( \frac{q^{2}}{Mr}\right) ^{a}}\right] ^{4/a}}{1-\frac{2M}{r}%
\left[ \frac{1}{1+\gamma \left( \frac{q^{2}}{Mr}\right) ^{a}}\right] ^{3/a}+%
\frac{q^{2}}{r^{2}}\left[ \frac{1}{1+\gamma \left( \frac{q^{2}}{Mr}\right)
^{a}}\right] ^{4/a}},  \tag{26}
\end{equation}

\begin{equation}
U^{003}=\frac{2z}{r^{2}}\frac{\frac{2M}{r}\left[ \frac{1}{1+\gamma \left(
\frac{q^{2}}{Mr}\right) ^{a}}\right] ^{3/a}-\frac{q^{2}}{r^{2}}\left[ \frac{1%
}{1+\gamma \left( \frac{q^{2}}{Mr}\right) ^{a}}\right] ^{4/a}}{1-\frac{2M}{r}%
\left[ \frac{1}{1+\gamma \left( \frac{q^{2}}{Mr}\right) ^{a}}\right] ^{3/a}+%
\frac{q^{2}}{r^{2}}\left[ \frac{1}{1+\gamma \left( \frac{q^{2}}{Mr}\right)
^{a}}\right] ^{4/a}}.  \tag{27}
\end{equation}

Using (25)-(27) and (24) we obtain the energy

\begin{equation}
E_{LL}=\frac{r}{2}\frac{\frac{2M}{r}\left[ \frac{1}{1+\gamma \left( \frac{%
q^{2}}{Mr}\right) ^{a}}\right] ^{3/a}-\frac{q^{2}}{r^{2}}\left[ \frac{1}{%
1+\gamma \left( \frac{q^{2}}{Mr}\right) ^{a}}\right] ^{4/a}}{1-\frac{2M}{r}%
\left[ \frac{1}{1+\gamma \left( \frac{q^{2}}{Mr}\right) ^{a}}\right] ^{3/a}+%
\frac{q^{2}}{r^{2}}\left[ \frac{1}{1+\gamma \left( \frac{q^{2}}{Mr}\right)
^{a}}\right] ^{4/a}}.  \tag{28}
\end{equation}

The energy in the Landau-Lifshitz prescription is plotted in Fig. 3 for $a=2$
, $\gamma =\frac{4}{9}$, $M=1$ and $q=0.1$.

\begin{figure}[H]
\begin{center}
\includegraphics[scale=1.2]{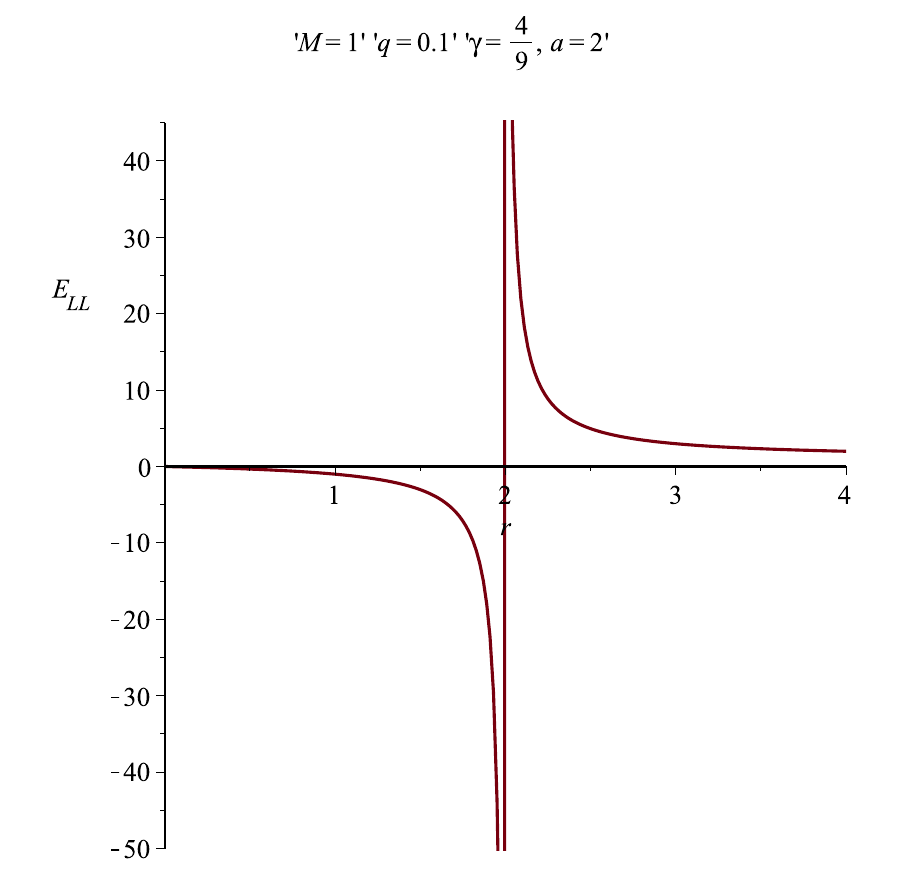}
\end{center}
\caption{Landau-Lifshitz energy versus the radial distance $r$.}
\label{f3}
\end{figure}

In Fig. 4 we present the graph of the energy near the origin for the same
values of $a$, $M$, $q$ and $\gamma $.

\begin{figure}[H]
\begin{center}
\includegraphics[scale=1.2]{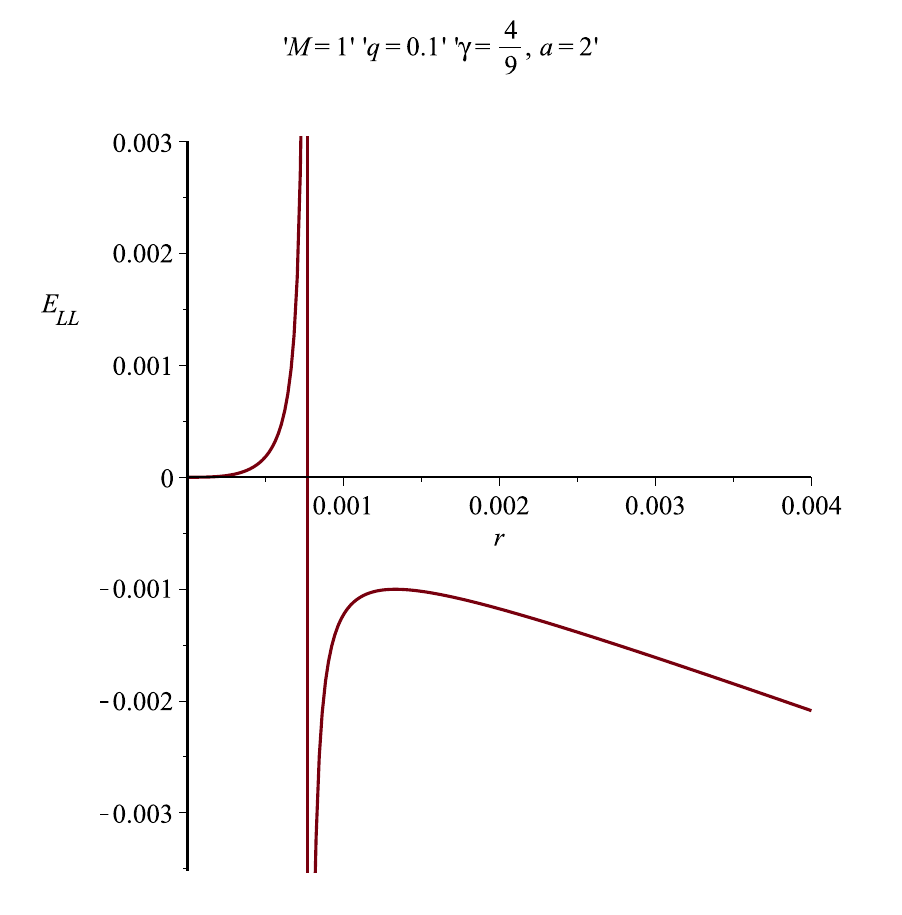}
\end{center}
\caption{Landau-Lifshitz energy versus the radial distance $r$ near the origin.}
\label{f4}
\end{figure}

\section{Weinberg Prescription and the Energy Distribution of the
Asymptotically Reissner-Nordstr\"{o}m Non-singular Black Hole}

The Weinberg energy-momentum complex [15] is given by the expression
\begin{equation}
W^{\mu \nu }=\frac{1}{16\pi }D_{,\,\lambda }^{\lambda \mu \nu },  \tag{29}
\end{equation}%
where the corresponding superpotentials are
\begin{equation}
D^{\lambda \mu \nu }=\frac{\partial h_{\kappa }^{\kappa }}{\partial
x_{\lambda }}\eta ^{\mu \nu }-\frac{\partial h_{\kappa }^{\kappa }}{\partial
x_{\mu }}\eta ^{\lambda \nu }-\frac{\partial h^{\kappa \lambda }}{\partial
x^{\kappa }}\eta ^{\mu \nu }+\frac{\partial h^{\kappa \mu }}{\partial
x^{\kappa }}\eta ^{\lambda \nu }+\frac{\partial h^{\lambda \nu }}{\partial
x_{\mu }}-\frac{\partial h^{\mu \nu }}{\partial x_{\lambda }},  \tag{30}
\end{equation}%
and
\begin{equation}
h_{\mu \nu }=g_{\mu \nu }-\eta _{\mu \nu }.  \tag{31}
\end{equation}%
The $W^{00}$ and $W^{0i}$ components represent the energy and the momentum
densities, respectively. In the Weinberg prescription the local conservation
law is respected:
\begin{equation}
W_{,\,\nu }^{\mu \nu }=0.  \tag{32}
\end{equation}%
By integrating $W^{\mu \nu }$ over the 3-space, one gets the expression for
the energy-momentum
\begin{equation}
P^{\mu }=\iiint W^{\mu 0}\,dx^{1}dx^{2}dx^{3}.  \tag{33}
\end{equation}%
Using Gauss' theorem and integrating over the surface of a sphere of radius $%
r$, the energy-momentum distribution takes the form:
\begin{equation}
P^{\mu }=\frac{1}{16\pi }\iint D^{i0\mu }n_{i}dS.  \tag{34}
\end{equation}

The nonvanishing superpotential components are as follows:

\begin{equation}
D^{100}=\frac{2x}{r^{2}}\frac{\frac{2M}{r}\left[ \frac{1}{1+\gamma \left(
\frac{q^{2}}{Mr}\right) ^{a}}\right] ^{3/a}-\frac{q^{2}}{r^{2}}\left[ \frac{1%
}{1+\gamma \left( \frac{q^{2}}{Mr}\right) ^{a}}\right] ^{4/a}}{1-\frac{2M}{r}%
\left[ \frac{1}{1+\gamma \left( \frac{q^{2}}{Mr}\right) ^{a}}\right] ^{3/a}+%
\frac{q^{2}}{r^{2}}\left[ \frac{1}{1+\gamma \left( \frac{q^{2}}{Mr}\right)
^{a}}\right] ^{4/a}},  \tag{35}
\end{equation}

\begin{equation}
D^{200}=\frac{2y}{r^{2}}\frac{\frac{2M}{r}\left[ \frac{1}{1+\gamma \left(
\frac{q^{2}}{Mr}\right) ^{a}}\right] ^{3/a}-\frac{q^{2}}{r^{2}}\left[ \frac{1%
}{1+\gamma \left( \frac{q^{2}}{Mr}\right) ^{a}}\right] ^{4/a}}{1-\frac{2M}{r}%
\left[ \frac{1}{1+\gamma \left( \frac{q^{2}}{Mr}\right) ^{a}}\right] ^{3/a}+%
\frac{q^{2}}{r^{2}}\left[ \frac{1}{1+\gamma \left( \frac{q^{2}}{Mr}\right)
^{a}}\right] ^{4/a}},  \tag{36}
\end{equation}

\begin{equation}
D^{300}=\frac{2z}{r^{2}}\frac{\frac{2M}{r}\left[ \frac{1}{1+\gamma \left(
\frac{q^{2}}{Mr}\right) ^{a}}\right] ^{3/a}-\frac{q^{2}}{r^{2}}\left[ \frac{1%
}{1+\gamma \left( \frac{q^{2}}{Mr}\right) ^{a}}\right] ^{4/a}}{1-\frac{2M}{r}%
\left[ \frac{1}{1+\gamma \left( \frac{q^{2}}{Mr}\right) ^{a}}\right] ^{3/a}+%
\frac{q^{2}}{r^{2}}\left[ \frac{1}{1+\gamma \left( \frac{q^{2}}{Mr}\right)
^{a}}\right] ^{4/a}}.  \tag{37}
\end{equation}

Substituting these expressions into (34), we obtain for the energy
distribution inside a $2$-sphere of radius\ $r$ the expression

\begin{equation}
E_{W}=\frac{r}{2}\frac{\frac{2M}{r}\left[ \frac{1}{1+\gamma \left( \frac{%
q^{2}}{Mr}\right) ^{a}}\right] ^{3/a}-\frac{q^{2}}{r^{2}}\left[ \frac{1}{%
1+\gamma \left( \frac{q^{2}}{Mr}\right) ^{a}}\right] ^{4/a}}{1-\frac{2M}{r}%
\left[ \frac{1}{1+\gamma \left( \frac{q^{2}}{Mr}\right) ^{a}}\right] ^{3/a}+%
\frac{q^{2}}{r^{2}}\left[ \frac{1}{1+\gamma \left( \frac{q^{2}}{Mr}\right)
^{a}}\right] ^{4/a}}.  \tag{38}
\end{equation}

The energy is plotted in Fig. 5 for $a=2$ , $\gamma =\frac{4}{9}$, $M=1$ and
$q=0.1$.

\begin{figure}[H]
\begin{center}
\includegraphics[scale=1.0]{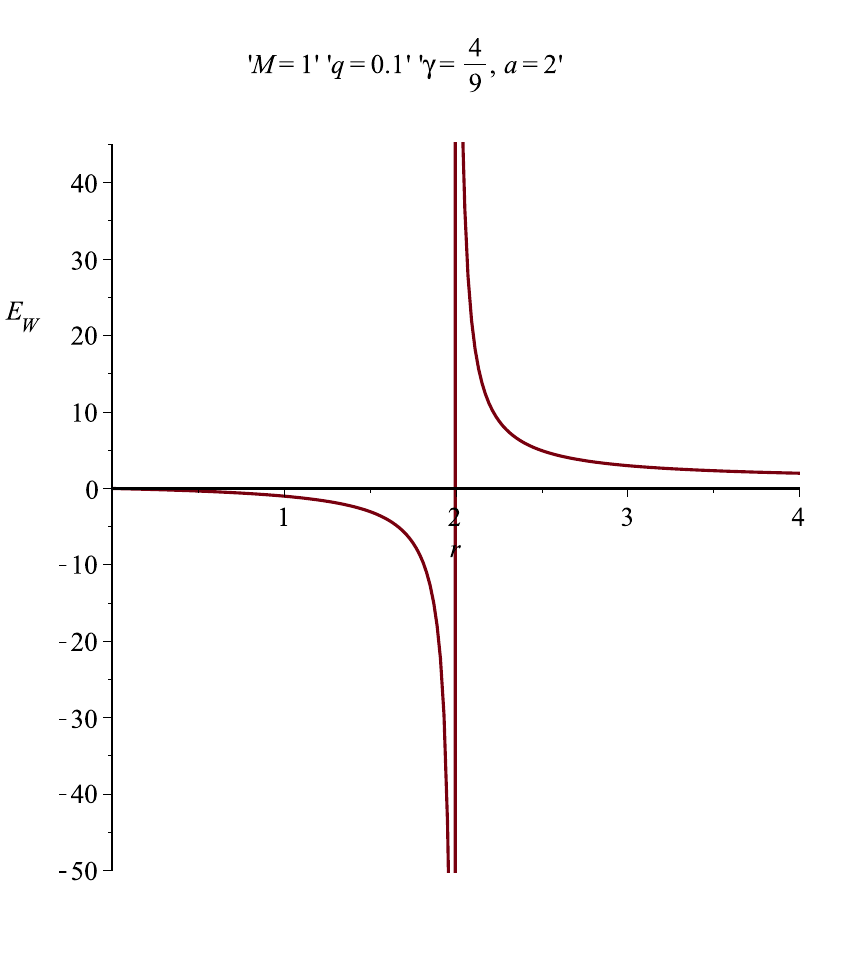}
\end{center}
\caption{Weinberg energy versus the radial distance $r$.}
\label{f5}
\end{figure}

Fig. 6 shows the behaviour of the energy near the origin for the same values
of $a$, $M$, $q$ and $\gamma $.

\begin{figure}[H]
\begin{center}
\includegraphics[scale=1.2]{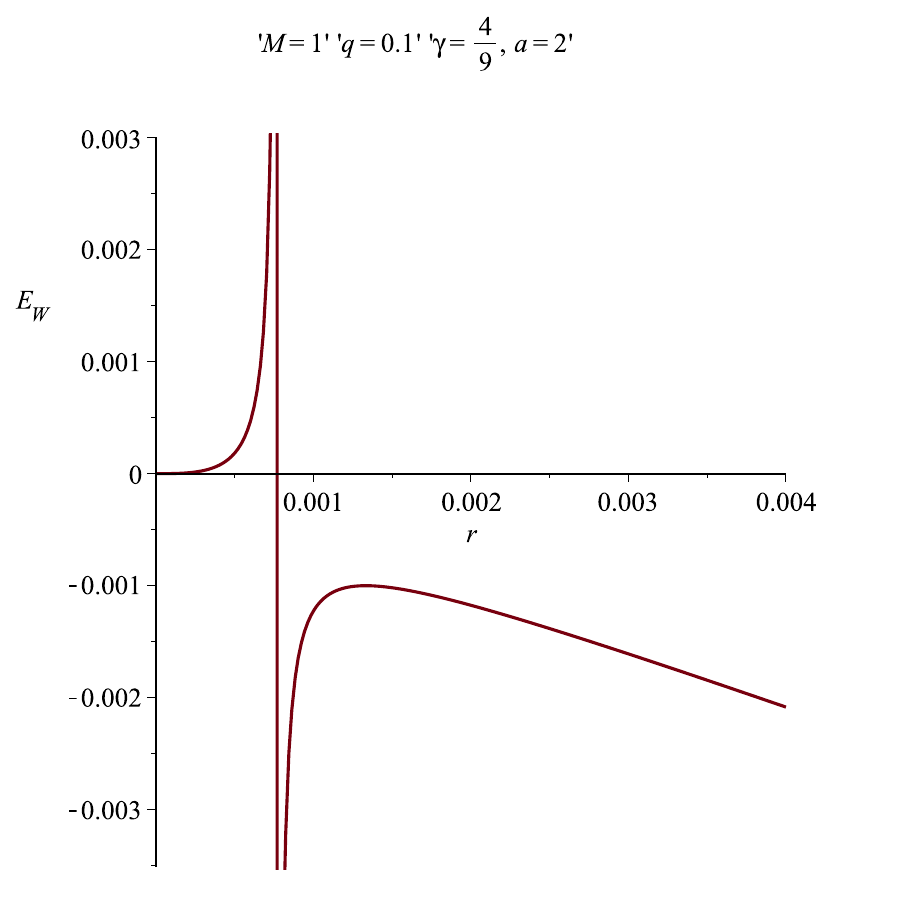}
\end{center}
\caption{Weinberg energy versus the radial distance $r$ near the origin.}
\label{f6}
\end{figure}

\section{M\o ller Prescription and the Energy Distribution of the
Asymptotically Reissner-Nordstr\"{o}m Non-singular Black Hole}

The expression for the M{\o }ller energy-momentum complex [14] is

\begin{equation}
\mathcal{J}_{\nu }^{\mu }=\frac{1}{8\pi }M_{\nu \,\,,\,\lambda }^{\mu
\lambda },  \tag{39}
\end{equation}%
where $M_{\nu }^{\mu \lambda }$ represent the M{\o }ller superpotentials:%
\begin{equation}
M_{\nu }^{\mu \lambda }=\sqrt{-g}\left( \frac{\partial g_{\nu \sigma }}{%
\partial x^{\kappa }}-\frac{\partial g_{\nu \kappa }}{\partial x^{\sigma }}%
\right) g^{\mu \kappa }g^{\lambda \sigma }.  \tag{40}
\end{equation}%
The M{\o }ller superpotentials $M_{\nu }^{\mu \lambda }$ are also
antisymmetric
\begin{equation}
M_{\nu }^{\mu \lambda }=-M_{\nu }^{\lambda \mu }.  \tag{41}
\end{equation}%
M{\o }ller's energy-momentum complex satifies the local conservation law%
\begin{equation}
\frac{\partial \mathcal{J}_{\nu }^{\mu }}{\partial x^{\mu }}=0,  \tag{42}
\end{equation}%
with $\mathcal{J}_{0}^{0}$ representing the energy density and $\mathcal{J}%
_{i}^{0}$ the momentum density components, respectively.

For the M\o ller prescription, the energy and momentum distributions are
given by
\begin{equation}
P_{\mu }=\iiint \mathcal{J}_{\mu }^{0}dx^{1}dx^{2}dx^{3}  \tag{43}
\end{equation}%
The energy distribution is calculated by using
\begin{equation}
E=\iiint \mathcal{J}_{0}^{0}dx^{1}dx^{2}dx^{3}.  \tag{44}
\end{equation}%
Applying Gauss' theorem one gets
\begin{equation}
P_{\mu }=\frac{1}{8\pi }\iint M_{\mu }^{0i}n_{i}dS.  \tag{45}
\end{equation}

In the M\o ller prescription we use Schwarzschild coordinates $\{t,$ $r,$ $%
\theta ,$ $\phi \}$ for the line element (4) and the metric function given
by eq. (1). We found that the only non-vanishing component of the M\o ller
superpotential (40) has the expression

\begin{equation}
\begin{split}
M_{0}^{01} &=\biggl\{\frac{2M}{r^2}\left[ \frac{1}{1+\gamma \left( \frac{q^{2}}{Mr}\right) ^{a}}%
\right] ^{3/a}\left[ 2-\frac{6\gamma \left( \frac{q^{2}}{Mr}\right) ^{a}}{%
1+\gamma \left( \frac{q^{2}}{Mr}\right) ^{a}}\right] - \\
&-\frac{2q^{2}}{r^{3}}\left[ \frac{1}{1+\gamma \left( \frac{q^{2}}{Mr}%
\right) ^{a}}\right] ^{4/a}\left[ 1-\frac{2\gamma \left( \frac{q^{2}}{Mr}%
\right) ^{a}}{1+\gamma \left( \frac{q^{2}}{Mr}\right) ^{a}}\right] \biggr\} \sin
\theta .
\end{split}
\tag{46}
\end{equation}

Combining eqs. (45) and (46) and after a few groupings of the terms, we
obtain the expression for the energy distribution:

\begin{equation}
\begin{split}
E_{M} &=M\left[ \frac{1}{1+\gamma \left( \frac{q^{2}}{Mr}\right) ^{a}}\right]
^{3/a}\left[ 1-\frac{3\gamma \left( \frac{q^{2}}{Mr}\right) ^{a}}{1+\gamma
\left( \frac{q^{2}}{Mr}\right) ^{a}}\right] - \\
&-\frac{q^{2}}{r}\left[ \frac{1}{1+\gamma \left( \frac{q^{2}}{Mr}\right) ^{a}%
}\right] ^{4/a}\left[ 1-\frac{2\gamma \left( \frac{q^{2}}{Mr}\right) ^{a}}{%
1+\gamma \left( \frac{q^{2}}{Mr}\right) ^{a}}\right] .
\end{split}
\tag{47}
\end{equation}

Furthermore, because of the vanishing of the spatial components of the M\o %
ller superpotential, all the momentum components are vanishing everywhere:

\begin{equation}
P_{r}=P_{\theta }=P_{\phi }=0.  \tag{48}
\end{equation}

In Fig. 7 we plot the energy in the M\o ller prescription for the parameter $%
a=2$ with $M=1$, $q=0.1$, and $\gamma =\frac{4}{9}$.

\begin{figure}[H]
\begin{center}
\includegraphics[scale=1.1]{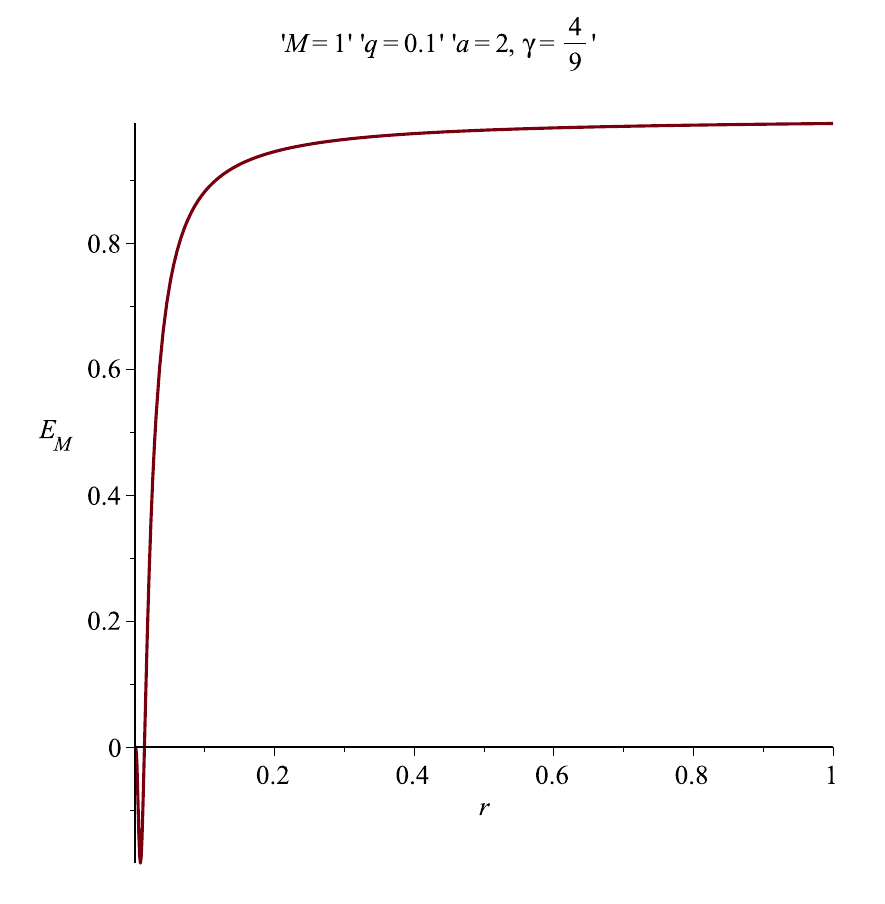}
\end{center}
\caption{M\o ller energy  versus the radial distance $r$.}
\label{f7}
\end{figure}

Fig.~8 shows the behaviour of the M\o ller energy near the origin for the
same values of $a$, $M$, $q$ and $\gamma $.

\begin{figure}[H]
\begin{center}
\includegraphics[scale=1.0]{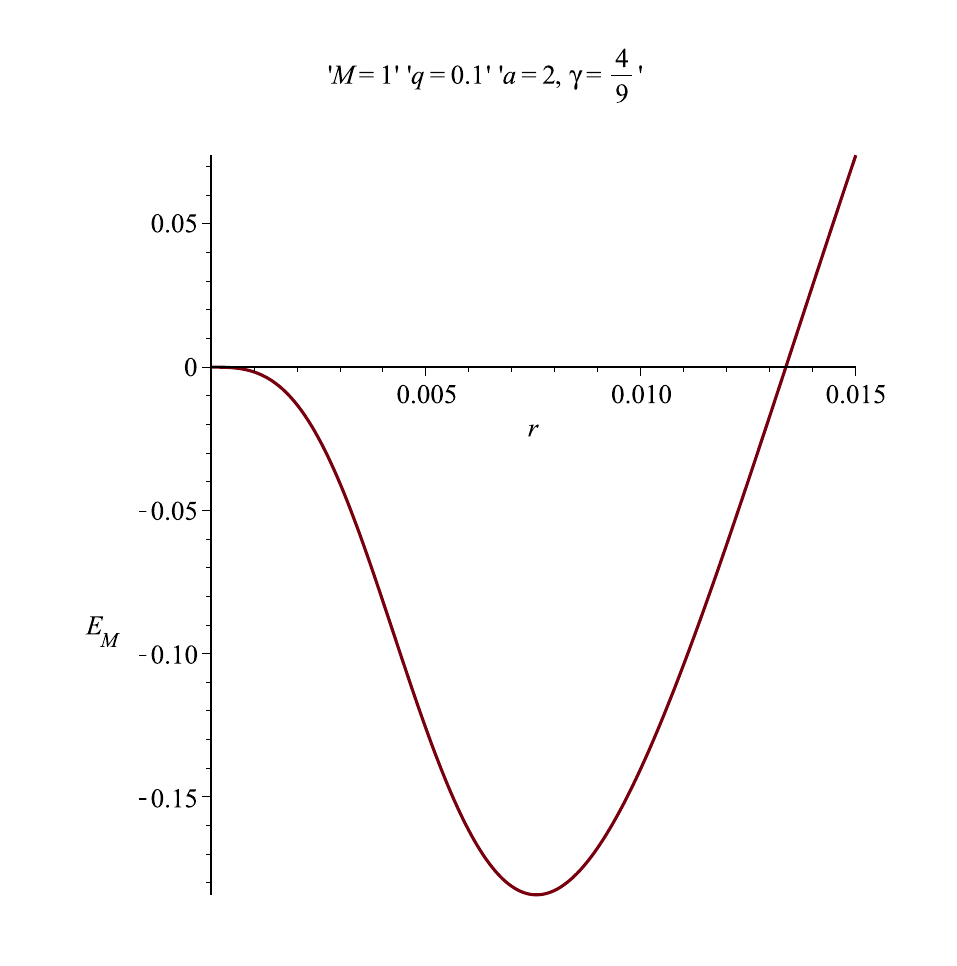}
\end{center}
\caption{M\o ller energy  versus the radial distance $r$ near the origin.}
\label{f8}
\end{figure}

In Fig. 9 we present a comparison of the energy distributions in the
Einstein, Landau-Lifshitz, Weinberg and M\o ller prescriptions for $a=2$, $%
M=1$, $q=0.1$ and $\gamma =\frac{4}{9}$.

\begin{figure}[H]
\begin{center}
\includegraphics[scale=0.7]{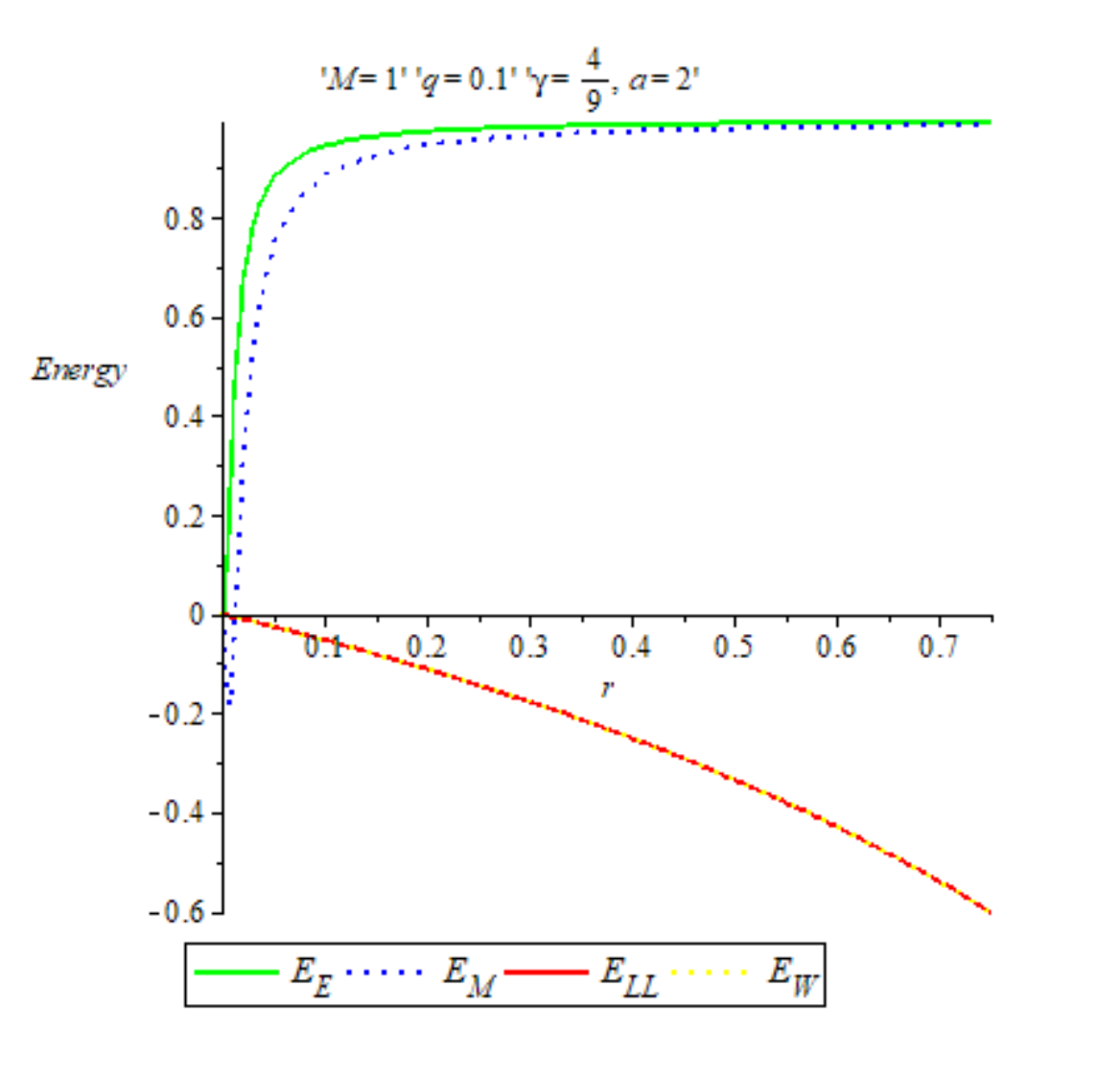}
\end{center}
\caption{Comparison of energy in the Einstein, Landau-Lifshitz, Weinberg, and M\o ller prescriptions versus the radial distance $r$.}
\label{f9}
\end{figure}

\section{Results and Discussion}

In this paper we have calculated the energy and momentum distributions for a
new spherically symmetric and charged asymptotically Reissner-Nordstr\"{o}m
non-singular black hole space-time geometry.

The Einstein, Landau-Lifshitz, Weinberg and M\o ller prescriptions have been
used, and we have found that these four prescriptions give the same result
regarding the momentum components, namely that all the momenta vanish. The
expressions of the energy are well-defined and physically meaningful showing
a dependence on the mass $M$, the charge $q,$ two parameters $\gamma $ and $a
$, and the radial coordinate $r$. Notice that the Landau-Lifshitz and
Weinberg energy-momentum complexes yield exactly the same expression for the
energy distribution. Also, for both presciptions the energy is equal to the
energy in the Einstein prescription divided by the term $1-\frac{2M}{r}\left[
\frac{1}{1+\gamma \left( \frac{q^{2}}{Mr}\right) ^{a}}\right] ^{3/a}+\frac{%
q^{2}}{r^{2}}\left[ \frac{1}{1+\gamma \left( \frac{q^{2}}{Mr}\right) ^{a}}%
\right] ^{4/a}$, in other words $E_{LL}=E_{W}=$ $\frac{E_{E}}{1-\frac{2M}{r}%
\left[ \frac{1}{1+\gamma \left( \frac{q^{2}}{Mr}\right) ^{a}}\right] ^{3/a}+%
\frac{q^{2}}{r^{2}}\left[ \frac{1}{1+\gamma \left( \frac{q^{2}}{Mr}\right)
^{a}}\right] ^{4/a}}$. Furthermore, the expressions of energy in the
Einstein and M\o ller prescriptions acquire the same value $M$, that is the
ADM mass, when $r\rightarrow \infty $ or for $q=0$. For $r\rightarrow \infty
$ the expression of energy is equal to the ADM mass also in the
Landau-Lifshitz and Weinberg prescriptions. This result is in agreement with
the result obtained by Virbhadra for the energy distribution of the
Schwarzschild black hole solution [27]. Indeed, in the particular case $q=0$%
, the energy in the Landau-Lifshitz and Weinberg prescriptions is given by \
the expression $\frac{M}{1-\frac{2M}{r}}$ that is obtained for the energy of
the Schwarzschild black hole solution in Schwarzschild Cartesian coordinates.

Table 1 shows the limiting behavior of the energy for $r\rightarrow 0$\ and $%
r\rightarrow \infty $, and in the particular case $q=0$.

\begin{table}[H]
\centering
\begin{tabular}{|c c c c|}
 \hline
Case & $r\rightarrow 0$ & $r\rightarrow \infty $ & $q=0$ \\
 \hline
Einstein & $0$ & $M$ & $M$ \\
 \hline
Landau-Lifshitz & $0$ & $M$ & $\frac{M}{1-\frac{2M}{r}}$ \\
 \hline
Weinberg & $0$ & $M$ & $\frac{M}{1-\frac{2M}{r}}$ \\
 \hline
M\o ller & $0$ & $M$ & $M$ \\
 \hline
\end{tabular}

\textbf{Table 1: Limiting behaviour}
\label{table:1}
\end{table}

For $r\rightarrow 0$ the new spherically symmetric and charged non-singular
black hole solution considered exhibits a de Sitter behaviour near zero, but
this does not necessarily imply that it satisfies the weak energy condition.

Fig. 10 shows the energy distributions in the Einstein, Landau-Lifshitz,
Weinberg and M\o ller prescriptions near the origin, as a function of $r$,
for the particular case $a=2$ and $\gamma =\frac{4}{9}$, with $M=1$, $q=0.1$.

\begin{figure}[]
\begin{center}
\includegraphics[scale=0.7]{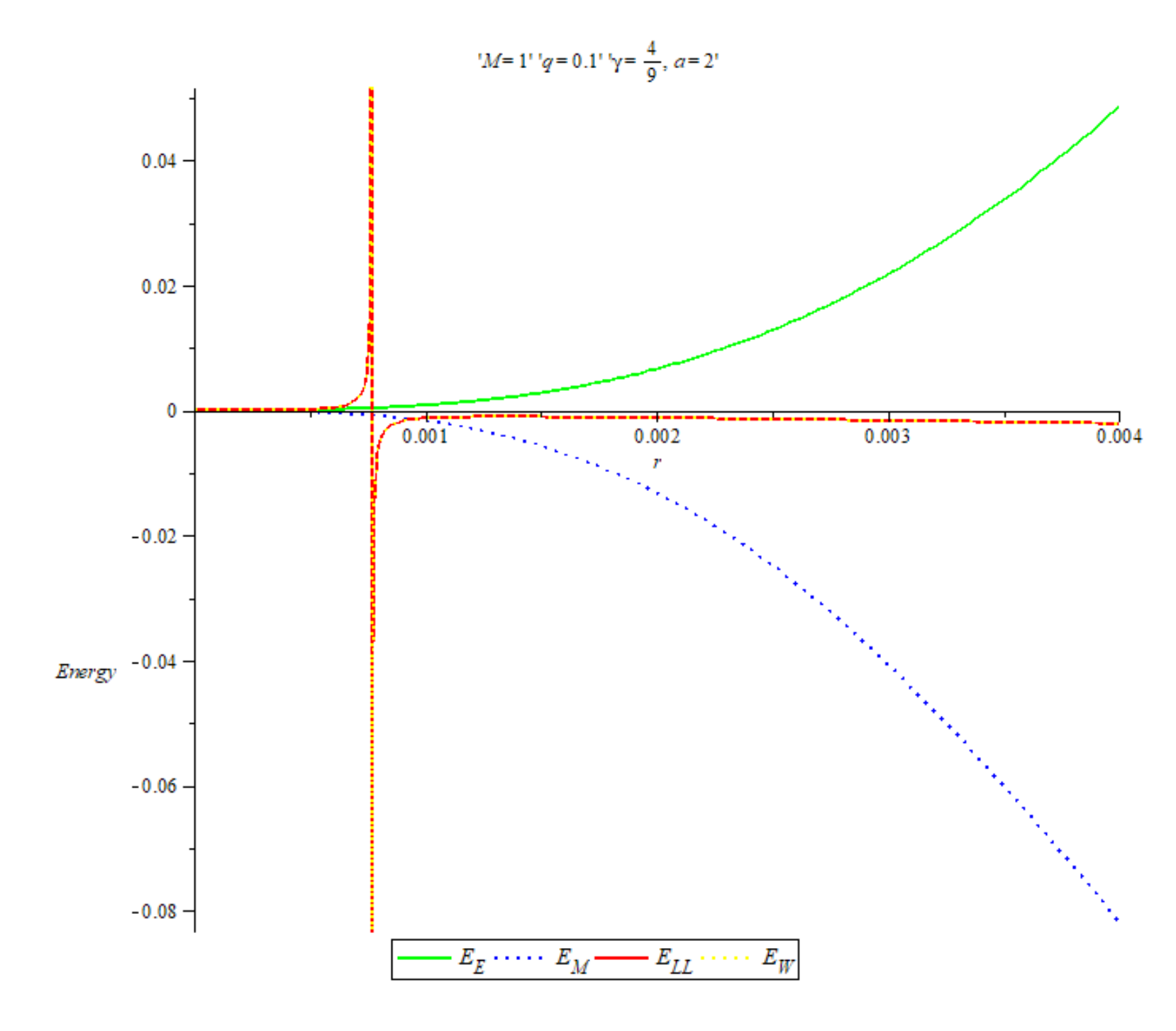}
\end{center}
\caption{Comparison of energy in the Einstein, Landau-Lifshitz, Weinberg, and M\o ller prescriptions versus the radial distance $r$ near the origin.}
\label{f10}
\end{figure}

The behaviour of the energy near the origin, that is for $r\rightarrow 0$,
is a special limiting case of particular interest. For some spacetime
geometries the metric ``goes infinite'' and, as a consequence, a singularity
appears, while the energy and momentum take extreme values. This particular
behavior is related to the specific nature of the space-time geometry. In
the case of the Einstein, Landau-Lifshitz, Weinberg and M\o ller
prescriptions, we notice that for $r\rightarrow 0$ the energy tends to zero.
Carrying out a more detailed investigation of the behavior of energy near
the origin we found that the Einstein energy tends to zero from positive
values, as expected from eq. (18), being an increasing function of $r$ which
tends to zero from the maximum value $M$ that is the ADM mass. The
Landau-Lifshitz and Weinberg energy presents points of divergence and takes
both positive and negative values. Analyzing Fig. 11, we deduce that for $r\lessapprox 0.00077$ the energy in these two prescriptions becomes positive
and tends to zero. Also, for $r\gtrapprox 1.995$ the Landau-Lifshitz and
Weinberg energy takes positive values and, finally, for large values of the
radial coordinate $r $, it becomes equal to the ADM mass $M$. The values of
the $r$ coordinate $r=0.00077$ and $r=1.995$ represent the two points of
divergence of the Landau-Lifshitz and Weinberg energy. The M\o ller energy
tends also to zero and, according to Fig.~12, close enough to zero, i.e. for
$r\lessapprox 0.0135$, it acquires negative values. For values of $r$
greater than $0.0135$, the M\o ller energy is positive and an increasing
function of $r$ acquiring the value $M$, which is the ADM mass, for $r\rightarrow \infty $.

\begin{figure}[H]
\begin{center}
\includegraphics[scale=1.1]{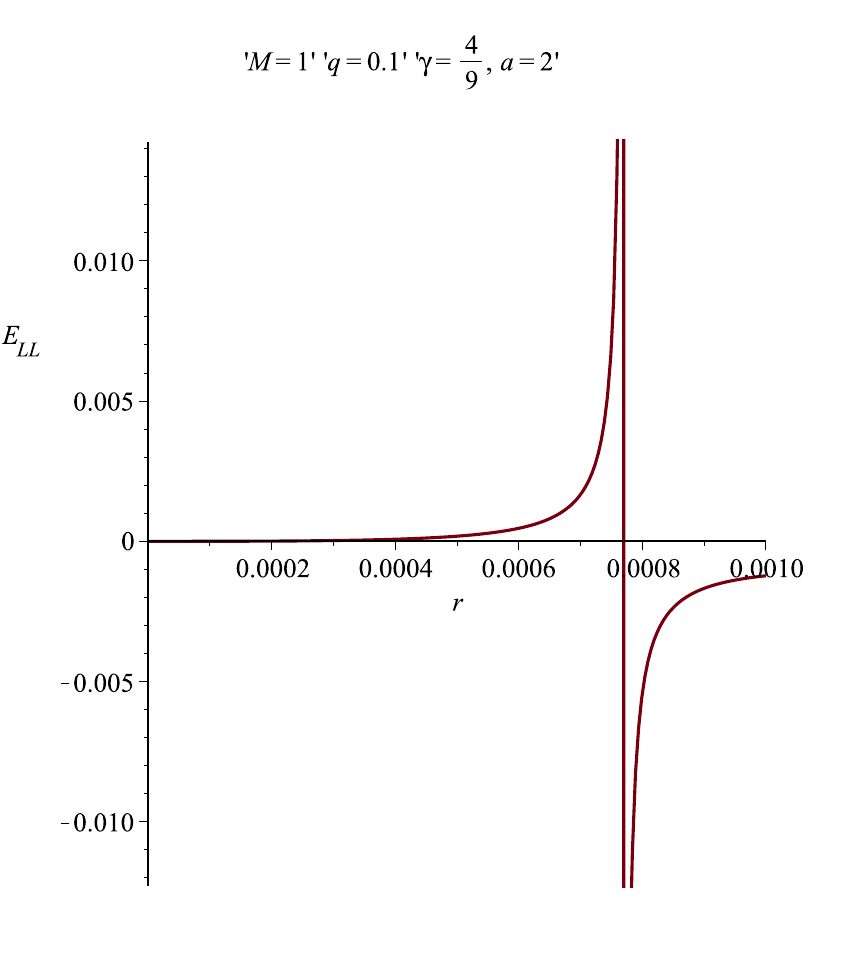}
\end{center}
\caption{Landau-Lifshitz energy versus the radial distance $r$ near the origin.}
\label{f11}
\end{figure}

\begin{figure}[H]
\begin{center}
\includegraphics[scale=1.1]{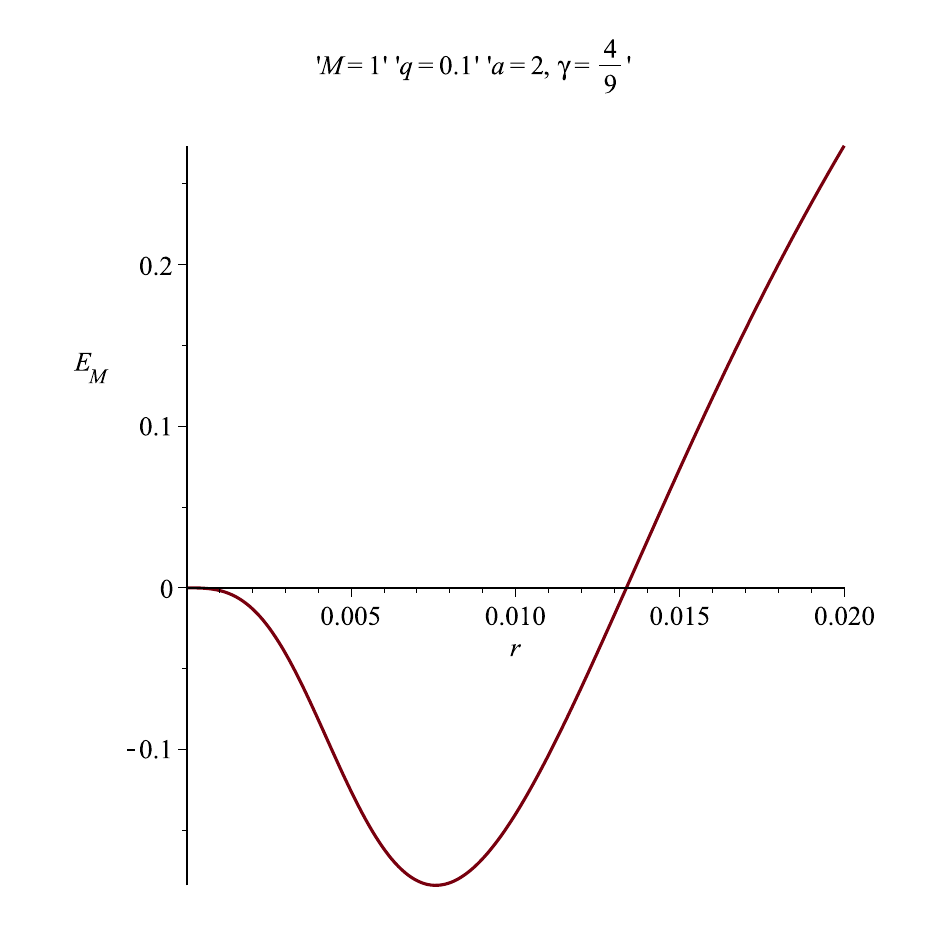}
\end{center}
\caption{M\o ller energy versus the radial distance $r$ near the origin.}
\label{f12}
\end{figure}

As we noticed, from Fig. 11, we observe that the Landau-Lifshitz and
Weinberg energy exhibits two points of divergence (singularities) whose
values depend on the values of the parameters $a$ and $\gamma $, and on the
mass $M$ and the charge $q$, jumping from negative to positive values, and
finally reaches the value of the ADM mass $M$ for $r\rightarrow \infty $.

A comment is useful concerning the divergence of the energy obtained in some of the prescriptions. Based on an idea presented in [57] in calculating conserved charges by the Komar formula in the context of TEGR, a regularization by means of a relocalization produced by properly modifying the Lagrangian of the gravitational field could possibly deal with this problem.

Regarding the physical significance of ``strange” parameters entering space-time metricsand leading to possibly ``strange” physical results like those in the present work, and despite the fact that in our case we have not taken up a further investigation in this direction, it is deemed proper to refer to some recent and interesting results obtained in the context of the extended general relativity known as $f(R)$ gravity and regarding a new exact solution describing a charged and spherically symmetric, black hole space-time that asymptotically behaves like an (A)dS space-time. In fact, the derived solution is an (A)dS Reissner-Nordström space-time depending on a positive parameter [58]. It is seen that this parameter is crucial for the thermodynamicsof this black hole. Remarkably, it is found that the entropy is not always proportional to the (outer) horizon areaand depends on the parameter values. Indeed, for a specific range of the parameter values, a non-trivial negative entropy emerges that can possibly point to forbidden regions where strange phase transitions may take place. Nevertheless, for parameter values restricted in the interval (0, 0.5) there are also a positive entropy, a positive Hawking temperature, a positive quasi-local energy, and a positive Gibbs free energy.

Finally, some very interesting results regarding the energy-momentum localization problem and its treatment by energy-momentum complexes are obtained in [59]. The authors investigate the gravitational energy localization in the context of both the GR and TEGR formalisms. For this analysis the complexes of Einstein, M\o ller, Landau-Lifshitz and Bergmann have been employed in order to evaluate the energy of the FLRW space-time. All the applied energy complexes give vanishing results at the background level. It must be stressed, that this result coincides with other previous results obtained by use of the Einstein and Landau-Lifshitz prescriptions. Furthermore, the authors calculated the gravitational energy of FLRW space-time at the level of cosmological linear perturbations up to first order and found that the expressions for the gravitational energy are non-vanishing, being identical in all the aforementioned prescriptions and related to the matter-energy density in the comoving gauge (except for the M\o ller prescription in GR) . The gravitational energy density for the universe filled with (a) non-relativistic matter, (b) radiation, (c) multiple scalar fields driving inflation, and (d) cosmological constant has also been calculated in this work.

These results come to support the use of the Einstein, Landau-Lifshitz,
Weinberg and M\o ller energy-momentum complexes for the evaluation of the
energy of a space-time geometry, while keeping in mind that the positive
energy region serves as a convergent lens and the negative one as a
divergent lens [60]. Also, the negativity of the expressions of energy in
the case of the Landau-Lifshitz, Weinberg and M\o ller prescriptions, for a
range of values of $a$, $\gamma $, $M$, $q$ and $r$, highlights the
existence of some difficulty in the physically meaningful interpretation of
the energy in certain regions of space-time.

\section{Conclusions}

The study of the asymptotically Reissner-Nordstr\"{o}m non-singular black
hole space-time geometry could be of great importance for black hole physics
as it would allow to test this particular black hole, the best approach
being gravitational lensing. In this light, the space-time described by the
metric given by eq. (4) with the metric function (1) allows obtaining useful
information concerning effects in strong gravitational lensing. As we noted
before, the positive and negative regions of the energy serve as convergent
and divergent lenses, [61]-[62] respectively, and the study of the behaviour
of the energy in the Einstein, Landau-Lifshitz, Weinberg and M\o ller
prescriptions near the event horizon also indicates what type of
microlensing can occur within each pseudotensorial prescription. The
behaviour of the energy near the event horizon could be analyzed by
performing a Taylor expansion of $E_{E}(r)$, $E_{LL}(r)$, $E_{W}(r)$ and $%
E_{M}(r)$ as a function of $r=0.00077$ and $r=1.995$ in the particular case
$a=2$, $\gamma =\frac{4}{9}$, $M=1$ and $q=0.1$. We choose these two values
for the radial coordinate $r$ because in the case of the event horizon the
equation $f(r)=0$, with $f(r)$ given by eq. (1) has two real roots, $%
r=0.00077$ and $r=1.995$, and, also, two complex roots. The energy in the
Einstein prescription near the event horizon $E_{HE}$ has only positive
values playing the role of a convergent lens. In the case of the
Landau-Lifshitz, Weinberg and M\o ller prescriptions, the expressions of the
energy near the event horizon $E_{HLL\text{ }}$, $E_{HW\text{ \ }}$and $%
E_{HM}$ take both positive and negative values and serve as a convergent and
divergent lens, respectively. To obtain more information about the
microlensing, a detailed analysis of the effects of dark energy on the
strong gravitational lensing in the case of the asymptotically
Reissner-Nordstr\"{o}m non-singular black hole is needed.

As a conclusion, the energy in the Einstein prescription takes only positive
values, while in the case of the\ Landau-Lifshitz, Weinberg and M\o ller
prescriptions the energy takes both positive and negative values depending
on the values of the radial coordinate $r$ and of $a$, $\gamma $, $M$ and $%
q. $ The apparent weakness of the Landau-Lifshitz, Weinberg and M\o ller
prescriptions could be justified by the properties of the particular metric
describing the asymptotically Reissner-Nordstr\"{o}m non-singular black
hole. Note that a similar behaviour of the M\o ller energy-momentum complex
was described in [63] and [64]. In the case of the asymptotically
Reissner-Nordstr\"{o}m non-singular black hole under study, this strange
behaviour is due, as in the case of the metrics presented in [60] and [61],
to the special properties of these black hole solutions that originate in
the coupling of the gravitational field to non-linear electrodynamics.

The pseudotensorial prescriptions used in this work constitute instructive
and useful tools for the energy-momentum localization. In this context, there exist also interesting approaches generalizing energy-momentum complexes to extended theories of gravity, where the Lagrangian depends on higher-order (up to n-th order) derivatives of the metric tensor. In [65], such a construction is shown to yield an affine, but non-covariant tensor as a generalization of the Landau-Lifshitz pseudo-tensor. For instance, an advantage of this approach lies in the fact that, in the weak-field limit and after a proper gauge choice is made, the energy-momentum pseudo-tensor after taking its average over a suitable region of space-time, is more tractable in order to calculate the energy. We consider a
challenging future issue to employ other pseudotensorial prescriptions, as
well as the teleparallel equivalent theory of general relativity, for
further investigation of the energy-momentum localization in the context of
the asymptotically Reissner-Nordstr\"{o}m non-singular black hole.\\\\
\end{document}